\documentclass{article}
\usepackage{spconf,amsmath,graphicx, amsfonts, url}
\usepackage{ctable}
\usepackage{tabularx,threeparttable}
\usepackage{kotex}
\usepackage{amssymb}
\usepackage{pifont}
\newcommand{\cmark}{\ding{51}}%
\newcommand{\xmark}{\ding{55}}%

\newcolumntype{Y}{>{\centering\arraybackslash}X}

\title{Gated Low-rank Adaptation for personalized Code-Switching Automatic Speech Recognition on the low-spec devices}
\name{Gwantae Kim$^{1}$, Bokyeung Lee$^{1}$, Donghyeon Kim$^{1}$ and Hanseok Ko$^{1}$}
\address{$^{1}$School of Electrical Engineering, Korea University, Seoul, South Korea}

\begin{document}
\ninept
\maketitle
\begin{abstract}
In recent times, there has been a growing interest in utilizing personalized large models on low-spec devices, such as mobile and CPU-only devices. However, utilizing a personalized large model in the on-device is inefficient, and sometimes limited due to computational cost. To tackle the problem, this paper presents the weights separation method to minimize on-device model weights using parameter-efficient fine-tuning methods.
Moreover, some people speak multiple languages in an utterance, as known as code-switching, the personalized ASR model is necessary to address such cases. However, current multilingual speech recognition models are limited to recognizing a single language within each utterance. To tackle this problem, we propose code-switching speech recognition models that incorporate fine-tuned monolingual and multilingual speech recognition models. Additionally, we introduce a gated low-rank adaptation(GLoRA) for parameter-efficient fine-tuning with minimal performance degradation. Our experiments, conducted on Korean-English code-switching datasets, demonstrate that fine-tuning speech recognition models for code-switching surpasses the performance of traditional code-switching speech recognition models trained from scratch. Furthermore, GLoRA enhances parameter-efficient fine-tuning performance compared to conventional LoRA.
\end{abstract}

\section{Introduction}
\label{sec:intro}
{\let\thefootnote\relax\footnotetext{This work was supported by the National Research Foundation of Korea(NRF) grant funded by the Korea government(MSIT) (NRF-2023R1A2C2005916). The authors were supported by Brand Engagement Networks. Corresponding Author: Hanseok Ko.}}
In recent times, there has been a growing interest in utilizing personalized large models on low-spec devices, such as mobile and CPU-only devices. To inject personal information into large models, fine-tuning using individual data is crucial. However, there are some challenges to using fine-tuned large models on small devices: 1. too much computational cost during the training phase. 2. one fine-tuned large model should exist per one personalized device. 3. too many memory resources for saving and processing whole fine-tuned large models on low-spec devices. Since full fine-tuning large models is hard even in high-spec devices, some parameter-efficient fine-tuning(PEFT) methods, such as Adapter\cite{houlsby2019parameter}, Prefix\cite{li2021prefix}, and LoRA\cite{hu2022lora}, have been proposed to address the problem. These methods give the effect of fine-tuning with a small amount of additional weights, and without changing the weights of the original large model. In personalized, on-device environments, these properties have the advantage of low computational cost during the training phase and separating large model weights and personalized model weights. In this case, the large model weights are stored in high-spec servers and the personalized weights are stored in low-spec devices, as described in Fig. 1.

\begin{figure}[t] 
\begin{center}
\includegraphics[width=1.0\linewidth]{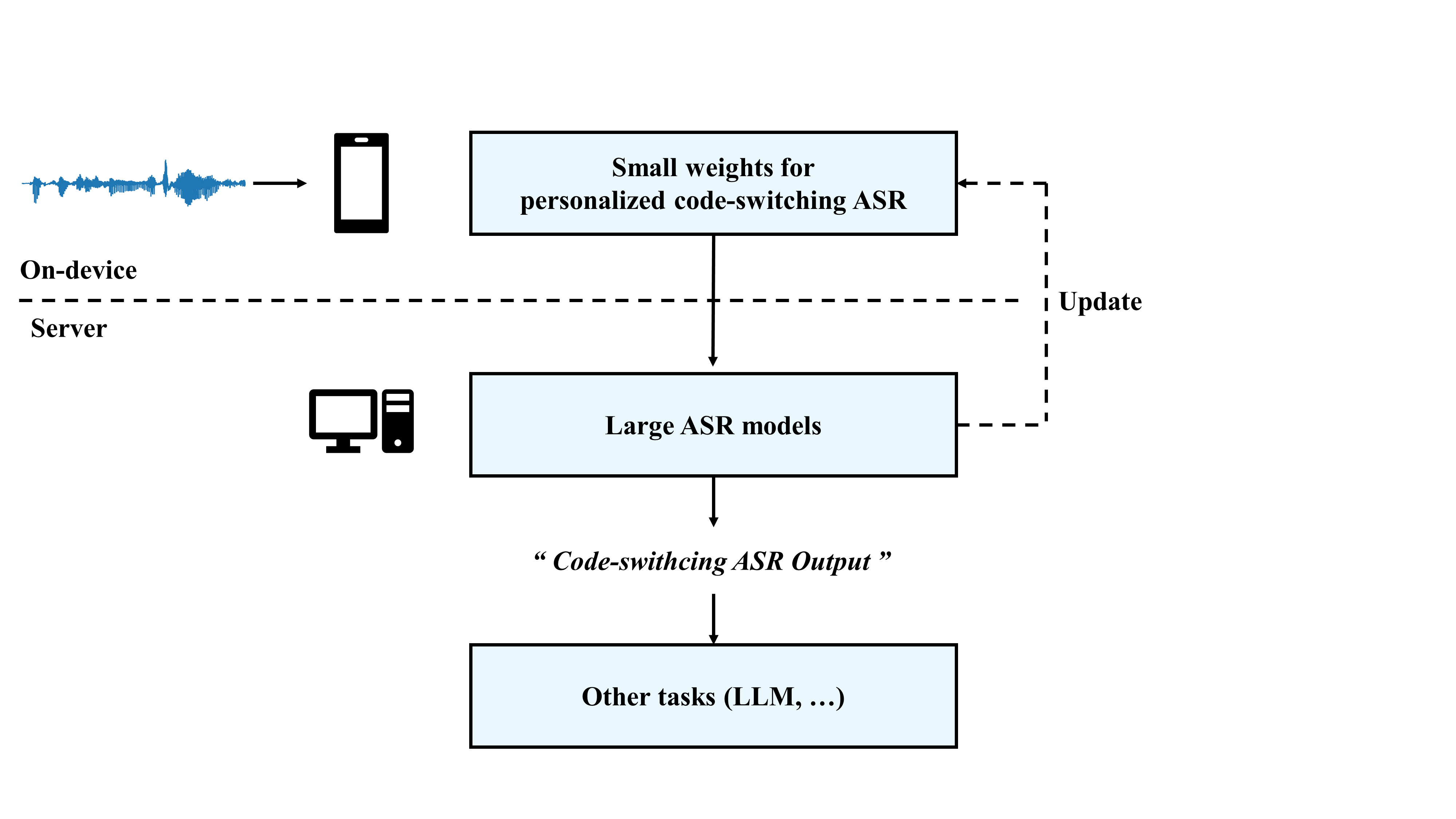}
\end{center}
\caption{Schema of the separating personalized weights and large model weights.}
\end{figure}

Automatic speech recognition(ASR) is one of the most frequently used AI modules on mobile devices. Monolingual ASR performs well in ideal recording environments\cite{kao2020orthogonal, panayotov2015librispeech, park2006achieving, gulati2020conformer, kim2005bayesian,graves2012sequence} and has already found applications\cite{park2011relationship, lee2007effective, beh2006dual} in various domains, including smartphones, human-computer interaction, robots, and more. As shown in Fig. 2, the attention to multilingual ASR is gradually increased by well-formulated data sets and models. The large speech corpora, such as Commonvoice\cite{ardila2020common} and MLS\cite{pratap2020mls}, are collected and distributed in public. Moreover, large-scale speech models, such as Wav2Vec2\cite{baevski2020wav2vec} and Whisper\cite{radford2023robust}, have enough capacity to address such large corpora. However, most previous multilingual ASR models recognize only one language per utterance. Since some people speak multiple languages in an utterance, as known as code-switching, the personalized ASR model is necessary to address such cases.

Moreover, the code-switching ASR assists in combining with multilingual LLMs. The multilingual ASR results can be used to the input of multilingual large language models(LLMs) when designing the speech-driven interaction agent. Since the tokenizer size of the LLMs is limited, unnecessary tokens should not be included in the tokenizer. However, the tokenizer should have the words with the same meaning not only in English but also in other languages if ASR cannot code-switch. For example, the multilingual utterance "I go to 경복궁" is recognized by ASR as "I go to Gyeongboggung" in English and "아이 고우 투 경복궁" in Korean without code-switching. Despite only one token "경복궁" is needed to understand "경복궁" in the multilingual LLMs, the token of English pronunciation "Gyeongboggung" should be added to the tokenizer without code-switching. Since "경복궁" and "Gyeongboggung" have the same meaning, adding both tokens to the tokenizer is inefficient. Therefore, code-switching ASR assists in reducing the tokenizer size of the LLMs.

\begin{figure}[t] 
\begin{center}
\includegraphics[width=1.0\linewidth]{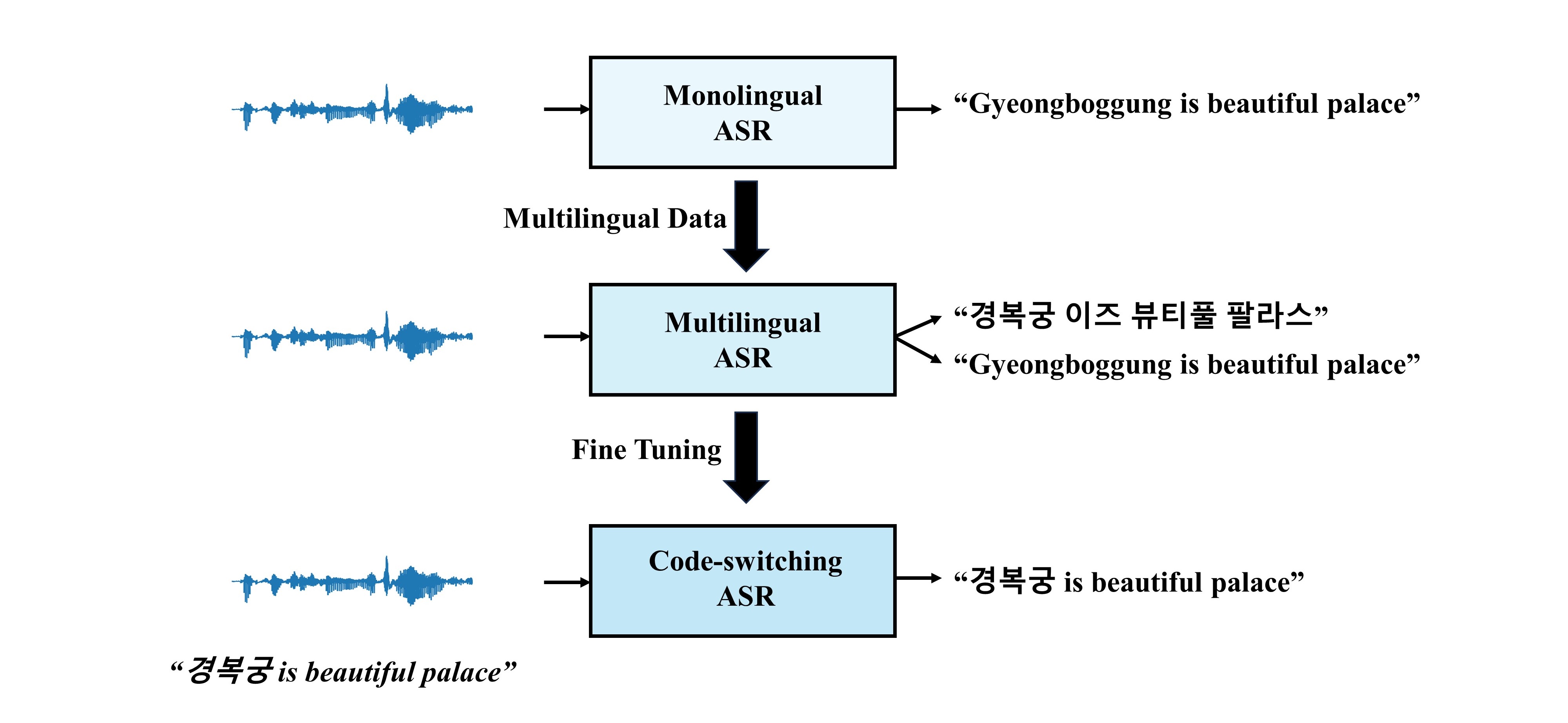}
\end{center}
\caption{Difference of the monolingual, multilingual, and code-switching ASR models.}
\end{figure}

Code-switching for multilingual ASR has a challenge that occurs with the same pronunciation between different languages. To tackle the problem, multilingual ASR methods\cite{babu2021xls, radford2023robust} first select or classify a language and recognize the transcription of the selected language only. Although this method can reduce confusion by pronunciation, the result is in one language, as mentioned above. Another approach\cite{anderson1994use, yi2021transfer} used the International Phonetic Alphabet (IPA) to address phonemes with similar pronunciation. Since this approach needs an additional IPA-to-language process, it may increase memory usage and latency. We tackle this problem by fine-tuning monolingual and multilingual ASR models using code-switching ASR data. The model is not limited to certain languages and can select any tokens in a dictionary.

In this paper, we propose a personalized code-switching ASR model with parameter-efficient fine-tuning large monolingual and multilingual ASR models to isolate personalized weights on the low-spec device and overcome the limitations of multilingual ASR models. We introduce a LoRA-style parameter-efficient fine-tuning method, named gated low-rank adaptation(GLoRA). GLoRA, which adopts gated linear units, improves the code-switching ASR performance with minimal parameter increase. We evaluate the code-switching ability of ASR models in three error rates that reflect linguistic features. With a series of experiments using a large Korean-English code-switching dataset, the fine-tuned code-switching ASR models outperform traditional code-switching speech recognition models and monolingual and multilingual ASR models in code-switching ASR. Especially, the fine-tuned models with GLoRA achieve competitive results to the full fine-tuning model with a small number of on-device weights.

Our contributions can be summarized as follows:
\begin{itemize}
    \item  We propose a weight separation schema, which utilizes large model weights on the server with remaining minimal on-device model weights.
    \item We present Korean-English code-switching ASR models with parameter-efficient fine-tuning methods to address the personalized code-switching behaviors of the users.
    \item We propose gated low-rank adaptation(GLoRA) to improve parameter-efficient fine-tuning performance without a large number of parameter increases.
\end{itemize}

\section{Related work}
\subsection{Code-switching ASR}
Recently, code-switching ASR models address One-to-English code-switching, and one can be other languages, such as Korean\cite{lee2021phonetic, wang2019exploring}, Chinese\cite{shan2019investigating, li2019towards, lovenia2022ascend}, and Arabic\cite{hussein2023textual}. Lee et al.\cite{lee2021phonetic} proposed phonetic variation modeling and language model adaptation for Korean-English code-switching. Wang et al.\cite{wang2019exploring} analyze the Korean language system for improving Korean-English Code-switching ASR. Shan et al.\cite{shan2019investigating} and Li et al.\cite{li2019towards} address model structure aspects of the code-switching ASR, such as LAS\cite{chan2016listen} and CTC\cite{graves2006connectionist}. Hussein et al.\cite{hussein2023textual} propose a pipeline structure for generating code-switching sentences. In this paper, the backbone models of the code-switching ASR model structures are updated to new large ASR models, including Wav2Vec2\cite{baevski2020wav2vec} and Whisper\cite{radford2023robust}.

\subsection{Fine-tuning}
There are many fine-tuning methods, also known as transfer learning, such as full fine-tuning, Adapter\cite{houlsby2019parameter}, prefix tuning\cite{li2021prefix}, and LoRA\cite{hu2022lora}. The full fine-tuning method trains all pre-trained weights. Although it often yields good performance, it is computing-inefficient. As the backbone model size grows larger, the computing-inefficient problem becomes a critical issue. To address the problem, Adapter\cite{houlsby2019parameter}, prefix\cite{li2021prefix}, or LoRA\cite{hu2022lora} layers are added to the backbone model. The weights of additional layers are only updated and the weights of the backbone model are frozen in the training stage. In this paper, the full fine-tuning and LoRA are used to comparison, and a new LoRA style model, named GLoRA, is proposed.

\section{Proposed method}

\subsection{Weight Separation}
The proposed weight separation method is summarized as storing PEFT weights on the low-spec device and sending them to the high-spec device with the user speech. Since the weights of the PEFT demand small computation resources, they can be stored in low-spec devices. To avoid the model initialization time consumption, the pre-trained ASR model should be initialized with a PEFT structure placeholder, and plug the PEFT weights during inference. For the training, the PEFT weights are fine-tuned using the transmitted speech and sent to the low-spec device.

Although the proposed weight separation method has a limitation in that the low-spec device should be connected to a network server, it has the advantage of efficiency. Moreover, the proposed method has privacy advantages because the speech and personalized weights are not stored on the server. Furthermore, the proposed method has personalized ASR performance advantages compared to utilizing pre-trained ASR models, because personalized weights address the dialects or habits of the individual users.

\subsection{Model Architecture}
\noindent\textbf{Whisper}
Whisper\cite{radford2023robust} is the transformer-based multilingual automatic speech recognition model, which is trained by weak supervision with large-scale speech recognition data. We use Whisper-tiny(39M parameters) and Whisper-small(244M parameters) as the backbone model in the Whisper model family. 
The Whisper predicts subword-level tokens from the log-mel spectrogram. To preserve the token numbering of the backbone model, the existing tokenizer of the backbone model remains and adds unseen tokens to the tokenizer using a code-switching dataset. The settings to extract the log-mel spectrogram are the same as Whisper: sampling rate=16000Hz, window length=25ms, hop length=10ms, and the number of mel coefficients=80. The cross-entropy loss is used to fine-tune the Whisper-based code-switching ASR model.

\noindent\textbf{Wav2Vec2}
Wav2Vec2\cite{baevski2020wav2vec} is the speech representation model that contains CNN and transformer layers. Wav2Vec2 is composed of a multi-layer convolutional feature encoder, which finds latent speech representations, and a Transformer to build representations capturing information from the entire sequence. Unlike the Whisper, Wav2Vec2 predicts character-level tokens from the raw audio. Since the base language of the dataset used in this paper is Korean, we use models and tokenizers of the Wav2Vec2-large-xlsr-Korean\footnote{\url{https://huggingface.co/kresnik/wav2vec2-large-xlsr-korean}} as backbone models and tokenizers. We also add unseen tokens to the tokenizer using a code-switching dataset. All raw audio is re-sampled to a 16000Hz sampling rate. The CTC loss\cite{graves2006connectionist} is used to fine-tune the Wav2Vec2-based code-switching ASR model.

\subsection{Low Rank Adaptation}
Low-rank adaptation is an efficient fine-tuning method for updating weight matrices without changing the original backbone model. A pre-trained weight matrix $W_0 \in \mathbb{R}^{d\times k}$ is updated by gradients $\Delta W$ during full fine-tuning. In the inference stage, the hidden features $y$ are calculated from input features $x$ as 
\begin{equation}
    h = (W_0+\Delta W)x.
\end{equation}
After full fine-tuning, pre-trained weights $W_0$ are not saved anymore, and they are changed into $W_0 + \Delta W$. This induces excessive memory usage. Since gradients for all $W_0$ should be calculated during the full fine-tuning stage, much more GPU memory is necessary. Moreover, all of the weights $W_0+\Delta W$ are saved for only one downstream task. If the task is changed, new weights $W_0 + \Delta W_2 $ should be trained and stored. It is over-consumption and memory inefficient. To address the problem, LoRA constrains updating $W_0$ by representing the latter with a low-rank decomposition $W_0 + \Delta W = W_0 + BA$, where $A \in \mathbb{R}^{d\times r}$, $B \in \mathbb{R}^{r\times k}$, and rank $r < min(d, k)$. During training, $W_0$ is frozen and does not receive gradient updates, while $A$ and $B$ contain trainable parameters. Note both $W_0$ and $\Delta W = BA$ are multiplied with the input features $x$, and their respective output vectors are summed coordinate-wise. The LoRA-modified forward pass yields:
\begin{equation}
    h = (W_0+\Delta W)x = W_0 x + BAx.
\end{equation}
The LoRA is memory-efficient and promises competitive results for fine-tuning.

\subsection{Gated Low Rank Adaptation}
The key idea of LoRA is estimating $\Delta W$ using $A$, $B$ of minimal parameters, without changing $W_0$. We aim to apply some tricks for performance improvement without hurting the original LoRA structure like:
\begin{equation}
    h = (W_0+\Delta W)x = W_0 x + f(BA)x.
\end{equation}
where $f(*)$ denotes any performance-improving functions.
We added the gated linear units(GLU) into LoRA with various parameter paths to design the tricks, as shown in Fig. 3. In type 1, input features $x$ and pre-trained output features $h$ are refined by GLU, concatenated, and passed through LoRA layer. In type 2, the GLU layers are injected into the input and output features of the LoRA. In type 3, the mid-level features of the LoRA are refined by GLU. Type 1, 2 GLoRA apply GLU to $x$ and $h$ in order to refine frozen features before extracting LoRA features $\Delta W$. Type 1 uses both pre-features and post-features, and Type 2 uses pre-features Type 3 GLoRA applies GLU to middle-level features of the LoRA. In this case, GLU is performed like an activation function.

\begin{figure}[t] 
\begin{center}
\includegraphics[width=1.0\linewidth]{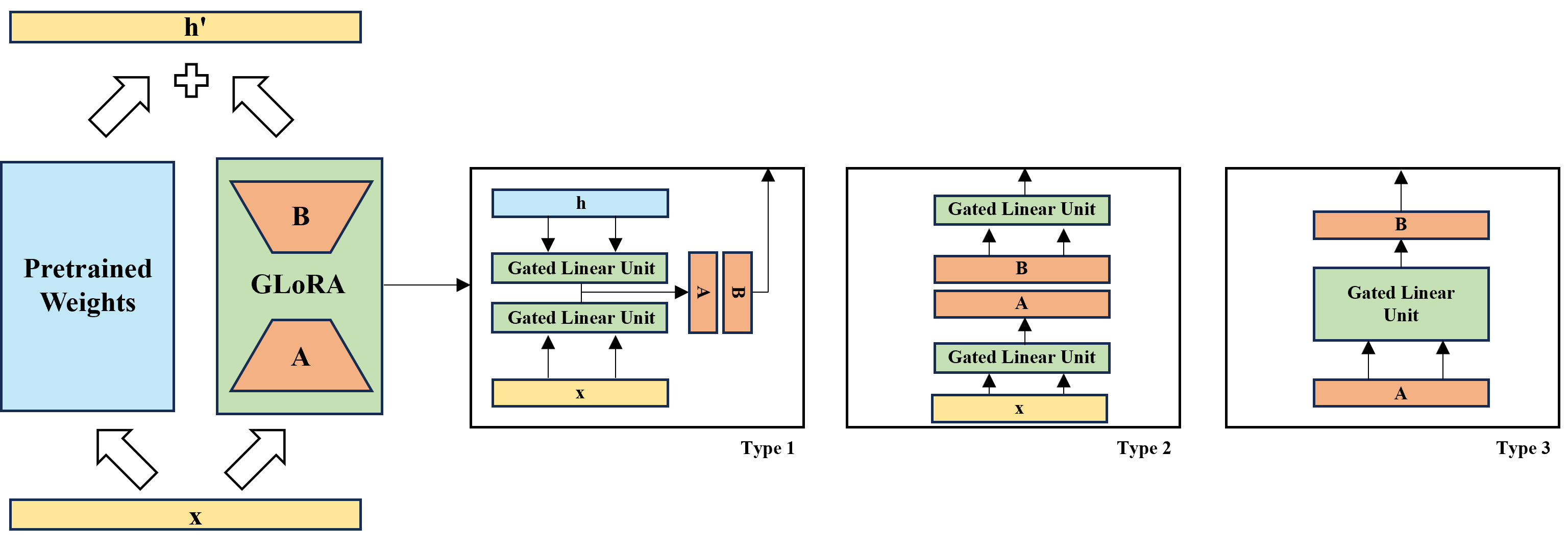}
\end{center}
\caption{Illustration of AuLoRA structures. Type 1: cross-attention. Type 2: self-attention. Type 3:GLU for hidden features. Type 4:GLU for low-rank features.}
\label{fig : AuLoRA}
\end{figure}

\section{Experiments}

\subsection{Datasets}
Experiments were performed on a Korean-English code-switching data set. The Korean-English code-switching data set, named KECS, is available in AI Hub\footnote{\url{https://aihub.or.kr/}}. The data set has 3502911 training utterances and 439607 test utterances with a 48kHz sampling rate. The waveform is resampled to follow the settings of the ASR models.
To evaluate Korean-English code-switching ASR models, we use word error rate(WER). Moreover, we adopt some additional measurements to consider linguistic differences. For the Korean-English case, character error rate(CER) and jamo error rate(JER) are calculated. In Korean, one character consists of many Jamos, and one word consists of many characters. As described in Fig 4, the Korean word "경복궁", which is pronounced "Gyeongboggung" in English, can be separated into characters or Jamos. Since Jamos and characters are related to pronunciation and words are related to meaning, CER and JER measure the phonetic recognition performance and WER measures the linguistic information recognition performance of the ASR model.

\begin{figure}[t] 
\begin{center}
\includegraphics[width=1.0\linewidth]{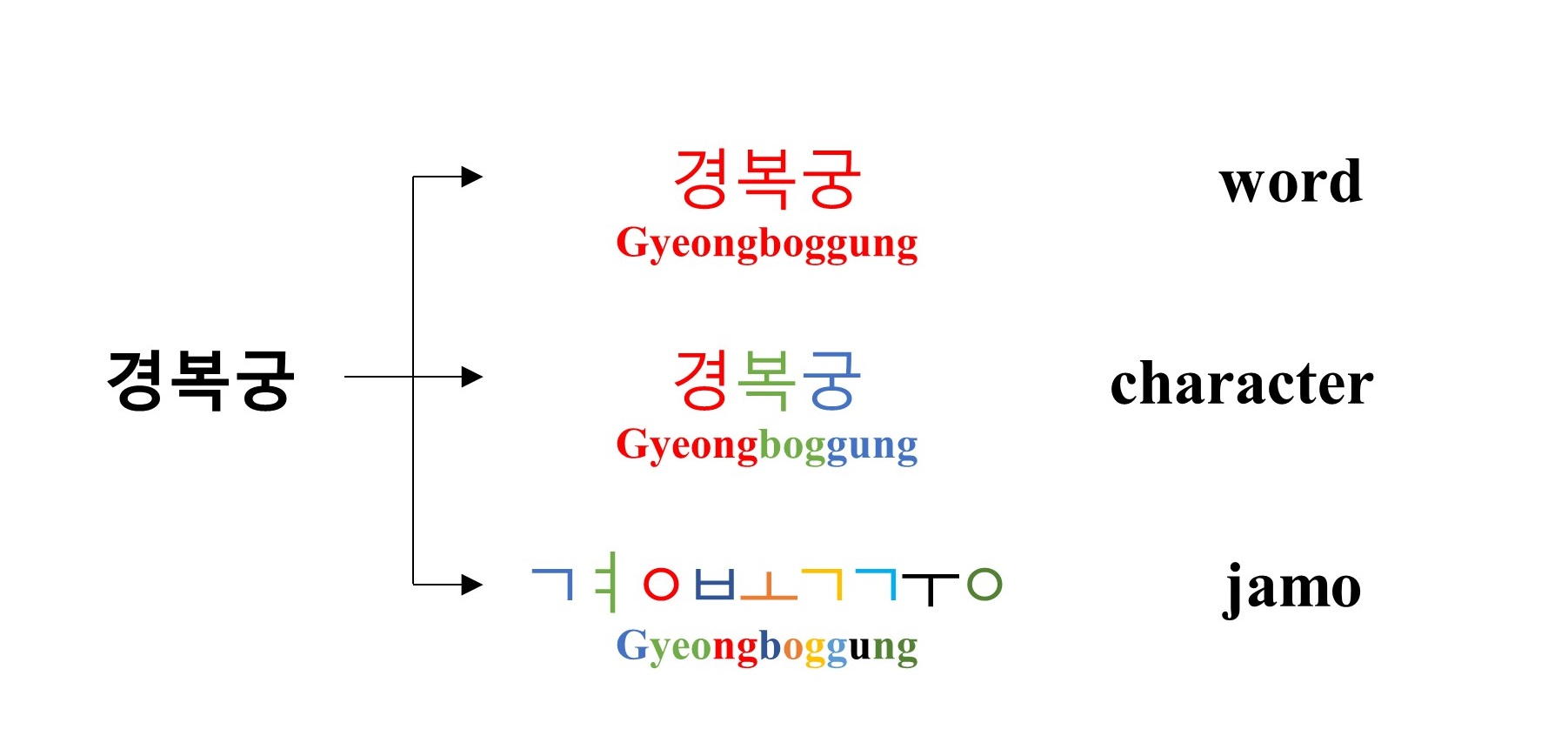}
\end{center}
\caption{Illustration of the Korean linguistic features. A Korean word can be separated into word, character, and Jamo.}
\label{fig : korean}
\end{figure}

\subsection{Evaluation Results and Discussion}

\begin{table*}[t]
\caption{The evaluation results of the proposed weight separation method using Korean-English code-switching ASR models.}
\centering
    \begin{tabular}{c|c|c|c|c|c|c}
    \specialrule{.2em}{.1em}{.1em}
    Model              & Fine-tuning &  LoRA   & WER$(\%)$ & CER$(\%)$ & JER$(\%)$ & trainable params. \\ \hline
    Conformer          &  -          &  \xmark & 82.098  & 75.151  & 63.582 & 118.8M \\ \hline
    Whisper-tiny       & \xmark      &  \xmark & 59.791  & 50.447  & 34.647 & 39M    \\
    Whisper-tiny       & \cmark      &  \xmark & \textbf{42.843}  & \textbf{21.309}  & \textbf{15.845} & 39M    \\
    Whisper-tiny       & \cmark      &  \cmark & 44.829  & 25.925  & 18.935 & 1.9M   \\ \hline
    Whisper-small      & \xmark      &  \xmark & 45.140  & 35.760  & 23.941 & 74M    \\
    Whisper-small      & \cmark      &  \xmark & 29.951  & 15.277  & 12.097 & 74M    \\
    Whisper-small      & \cmark      &  \cmark & \textbf{33.575}  & \textbf{17.598}  & \textbf{13.671} & 7.3M   \\ \hline
    Wav2Vec2-large(ko) & \xmark      &  \xmark & 72.360  & 55.965  & 38.371 & 313M   \\
    Wav2Vec2-large(ko) & \cmark      &  \xmark & 42.371  & 21.952  & 18.068 & 313M   \\
    Wav2Vec2-large(ko) & \cmark      &  \cmark & 55.141  & 55.394  & 38.235 & 7.1M   \\  
    \specialrule{.2em}{.1em}{.1em}
    \end{tabular}
\end{table*}

\begin{table}[t]
\caption{The evaluation results of the proposed GLoRA using Korean-English code-switching ASR models.}
\centering
    \begin{tabular}{c|c|c|c|c}
    \specialrule{.2em}{.1em}{.1em}
    Model          &  GLoRA & WER$(\%)$ & CER$(\%)$ & JER$(\%)$ \\ \hline
    Whisper-tiny   &  ori.  & 44.829  & 25.925  & 18.935\\
    Whisper-tiny   &  Type1 & 41.740  & 23.721  & 17.651\\
    Whisper-tiny   &  Type2 & 45.428  & 26.375  & 19.225\\
    Whisper-tiny   &  Type3 & 44.919  & 25.925  & 18.935\\ \hline
    Wav2Vec2       &  ori.  & 55.141  & 55.394  & 38.235\\
    Wav2Vec2       &  Type1 & 53.468  & 54.208  & 36.646\\
    Wav2Vec2       &  Type2 & 54.663 & 54.802  & 37.980 \\
    Wav2Vec2       &  Type3 & 54.757 & 54.813  & 38.136 \\
    \specialrule{.2em}{.1em}{.1em}
    \end{tabular}
\end{table}

The evaluation results of the Korean-English code-switching are summarized in Table 1. To evaluate the conventional on-device code-switching ASR model, the Conformer\cite{gulati2020conformer} model is trained from scratch using Espnet2\cite{watanabe2018espnet} first. The original and fine-tuned Whisper-tiny, Whisper-small, and Wav2Vec2-large models outperform the Conformer model by a large margin because the Conformer model is not pre-trained by other large corpora for ASR.

To evaluate performance improvement of the weight separation process, the fine-tuned whisper-tiny model (small on-device ASR model) and LoRA fine-tuned whisper-small model (proposed weight separation method with large ASR) are compared, which are marked bold in Table 1. Although the Whisper-small LoRA fine-tuned model has lower parameters, which is 7.3M on-device parameters, it outperforms the Whisper-tiny full fine-tuned model which has 39M on-device parameters. Therefore, the efficiency and ASR performance gap is promised between the on-device code-switching ASR and the proposed weight separation method. The results of the Wav2Vec2-large(Korean) model are also summarized in Table 1. Since the pre-trained model is trained with Korean corpora only, the results are worse than the multilingual pre-trained ASR model. Still, it shows a similar tendency.

On the same model size and server-connected network setting, a large fully fine-tuned model achieves better performance than the pre-trained model and LoRA fine-tuned model. However, training and saving fully fine-tuned large models for each user need too much memory. Furthermore, the performance gap between a fully fine-tuned model and a LoRA fine-tuned model is not significant. Therefore, storing LoRA parameters for each user, especially saving them on the user's device, is an efficient and secure way.

The evaluation results of the proposed GLoRA on Whisper-tiny and Wav2Vec2-large(ko) models are summarized in Table 2. The GLoRA-type1, which used not only pre-trained input features but also output features to extract LoRA features, outperforms the original and other types of GLoRA. The GLoRA-type2 and GLoRA-type3 achieve better performance on Wav2Vec2, but worse performance on Whisper-tiny, though the performance gap is not large.

\noindent\textbf{Token normalization} There are two types of Unicode representation for Korean Jamo tokens. The unnormalized tokenizer uses the U+11xx lineup and the normalized tokenizer uses the U+31xx lineup. The comparison is described in Table 3. The fine-tuned model trained by a normalized tokenizer is 0.8\% better than the unnormalized tokenizer in WER. We found that the tokenizer of the Whisper-small model has both types of tokens and the inference results of the original Whisper-small model always consist of normalized tokens. These results may infer that the Whisper-small model is trained by normalized tokens. Thus, the result that the fine-tuned model with normalized tokens is better than unnormalized tokens is reasonable.

\begin{table}[t]
\caption{The effect of language token normalization on Korean-English code-switching task. Backbone model: Whisper-small}
\label{table:ab1}
\centering
    \begin{tabular}{c|c|c|c}
    \specialrule{.2em}{.1em}{.1em}
    Lang. Tokenizer   & $WER (\%)$ & $CER (\%)$ & $JER (\%)$ \\ \hline
    Unnormalized      & 30.757  & 15.702  & 12.437 \\
    Normalized        & 29.951  & 15.277  & 12.097 \\
    \specialrule{.2em}{.1em}{.1em}
    \end{tabular}
\end{table}

\section{Conclusion}
In this study, we proposed a weight separation method for personalized code-switching ASR on low-spec devices. We also presented the gated linear unit-based LoRA structure, named GLoRA. The proposed weight separation method has the advantages of efficiency, privacy, and personalized ASR performance. The GLoRA, which used gated linear units and pre-trained output features to extract LoRA features, outperforms the original LoRA. Through the series of experiments, the proposed methods are validated on personalized code-switching ASR. In the future, the weight separation method will be applied to customer service-level applications and the GLoRA will be employed for other tasks that use large neural network models.

\bibliographystyle{IEEEbib}
{\footnotesize\bibliography{ref}}

\end{document}